\documentclass[12pt,preprint]{aastex}

\shorttitle{COOL VERSUS ULTRACOOL WHITE DWARFS}
\shortauthors{J. Farihi}

\begin{document}

\title{COOL VERSUS ULTRACOOL WHITE DWARFS}

\author{J. Farihi\altaffilmark{1,2}}

\altaffiltext{1}{Gemini Observatory,
			Northern Operations,
			670 North A'ohoku Place,
			Hilo, HI 96720; jfarihi@gemini.edu}
\altaffiltext{2}{Department of Physics \& Astronomy,
			University of California,
			Los Angeles, CA 90095}

\begin{abstract}

A preliminary $BVRIJHK$ analysis of the white dwarfs SSSPM
J2231$-$7514 and SSSPM J2231$-$7515 is presented.  Although
both stars were reported to have $T_{\rm eff}<4000$ K, the
analysis here indicates $T_{\rm eff}\approx4250$ K for both SSSPM
J2231$-$7514 and SSSPM J2231$-$7515.  Given substantial scientific
interest in the coolest extant degenerate stars, it is necessary to
distinguish sub 4000 K objects from the bulk of cool white dwarfs.
This analysis reiterates the importance of near infrared observations
in constraining the spectral energy distributions and effective
temperatures of the coolest white dwarfs and briefly discusses
their possible origins.

\end{abstract}


\section{INTRODUCTION}

The study of cool white dwarfs with $T_{\rm eff}>4000$ K
has been artfully mastered by P. Bergeron and collaborators.
They have shown that with $BVRIJHK$ photometry alone, the
effective temperature and atmospheric composition of cool
degenerates can be determined with a high degree of accuracy.
In addition, if the white dwarf has a known distance or Balmer
lines, then the surface gravity (hence mass and radius) can be
determined quite well.  Comparisons of predicted versus measured
absolute magnitudes and radii for white dwarfs with trigonometric
parallaxes have confirmed their findings \citep{ber97,leg98,ber01b}.

Known ultracool white dwarfs ($T_{\rm eff}<4000$ K) are
spectrally distinct objects and should be considered a separate
class of degenerate star.  The overall shape of their emergent
flux is strongly influenced by opacity due to collisions between
${\rm H_2}$ molecules in pure hydrogen atmospheres or between He
and ${\rm H_2}$ in mixed atmospheres (for a great review, see \S2
of \citealt{ber01a}).  This collision induced absorption (CIA) has
been observed to suppress flux at near infrared and red optical
wavelengths.  At present there exist only 4 known ultracool white
dwarfs with effective temperature estimates based on published
optical and near infrared data \citep{har99,har01,hod00,ber01a,
ber02,opp01a,far04}.

This paper presents an examination of a few ultracool white dwarf
candidates based on existing data.  A preliminary optical plus near
infrared spectral energy distribution analysis of the cool white dwarfs
SSSPM J2231$-$7514 and SSSPM J2231$-$7515 supports effective temperatures
near or above 4250 K and little, if any, flux suppression due to CIA.  A
brief examination of the data available on F351-50 indicates a possible
effective temperature above 4000 K as well.

\section{DATA \& ANALYSIS}

\citet{sch02} reported the discovery of a comoving pair of
faint high proper motion stars which were spectroscopically
determined to be cool DC white dwarfs.  These white dwarfs are
of interest  because they are likely to be within 20 pc of the sun
and potentially cooler than previously known degenerates at this
distance.

\subsection{Photometry}

Optical $BVRI$ photometric data were taken from \citet{sch02}.
The $BRI$ magnitudes for the white dwarfs are from the SuperCOSMOS 
Sky Survey (SSS) and photographic in nature, hence the uncertainties
are relatively large \citep{ham01a}.  These were converted to the
Johnson-Cousins system using the appropriate transformations
\citep{bla82,bes86,sal04}.

The 2MASS All Sky image
database \citep{cut03} shows both SSSPM J2231$-$7514 and
SSSPM J2231$-$7515 at positions $22^{\rm h} 30^{\rm m} 40.08^
{\rm s}$, $-75\arcdeg 13\arcmin 56.7''$ \& $22^{\rm h} 30^{\rm m}
33.63^{\rm s}$, $-75\arcdeg 15\arcmin 25.6''$ (J2000, epoch 2000
Oct 08) respectively at all three wavelengths.  A comparison of
the 2MASS $J$ band image with the digitized UKST $I$ band image
(epoch 1993) confirms the identity of the stars with the correct
published proper motion of $\mu=1.87'' \ {\rm yr}^{-1}$ at $\theta
=167.5\arcdeg$ \citep{sch02}.  $JHK_s$ magnitudes were extracted
from the 2MASS database at the above positions.  All data are
listed in Table \ref{tbl-1}.

\subsection{Colors \& Atmosphere}

The brighter and fainter binary components have $V-J$ colors
of 1.94 and 2.01 respectively.  This color index involves the
two filters with the smallest measurement errors and are therefore
the most reliable (especially compared to color indices involving
$BRI$).  In addition, their $V-K_s$ colors are 2.16 and  2.15
respectively, with slightly larger uncertainty in the $K_s$
magnitudes.  If accurate, these colors indicate that both stars
are very likely to have effective temperatures above 4000 K,
regardless of atmospheric composition.  In the following,
log $g=8.0$ is assumed.

White dwarfs with hydrogen atmospheres can possess near
infrared colors that are bluer than those stated above (due
to CIA), beginning at $T_{\rm eff}<5000$ K.  By 4000 K, their
colors will certainly be much bluer than those implied by Table
\ref{tbl-1} \citep{ber95,ber97,ber01a,ber01b}.  For cool white
dwarfs with normal mass (log $g\sim8.0$) in general, the predicted
and measured $V-J$ colors for hydrogen atmospheres do not become as
red as those associated with helium atmospheres.  For example, $V-J$
reaches a maximum around 1.9 for log $g=8.0$ and around 1.8 for log
$g=8.5$ in cool hydrogen atmosphere models for $T_{\rm eff}=4250$ K.
However, colors as red as $V-J\approx2.0$ have been observed and
associated with hydrogen rich atmospheres \citep{ber95,ber97,ber01b}.

Cool helium atmosphere white dwarfs are predicted and measured to
attain colors this red in $V-J$ around $T_{\rm eff}=4500$ K.  However,
the corresponding near infrared colors for helium atmospheres are also 
red, with $J-K\ga0.3$ corresponding to a $V-J\sim2.0$.  Thus, if the
2MASS photometry is accurate, SSSPM J2231$-$7514 and SSSPM J2231$-$7515
are likely to have hydrogen rich atmospheres, but a helium rich
composition cannot be ruled out.  In \S2.3, model fits using both
hydrogen and helium atmospheres are considered.  

\subsection{Spectral Energy Distributions \& Temperatures}
 
\begin{figure}

\plotone{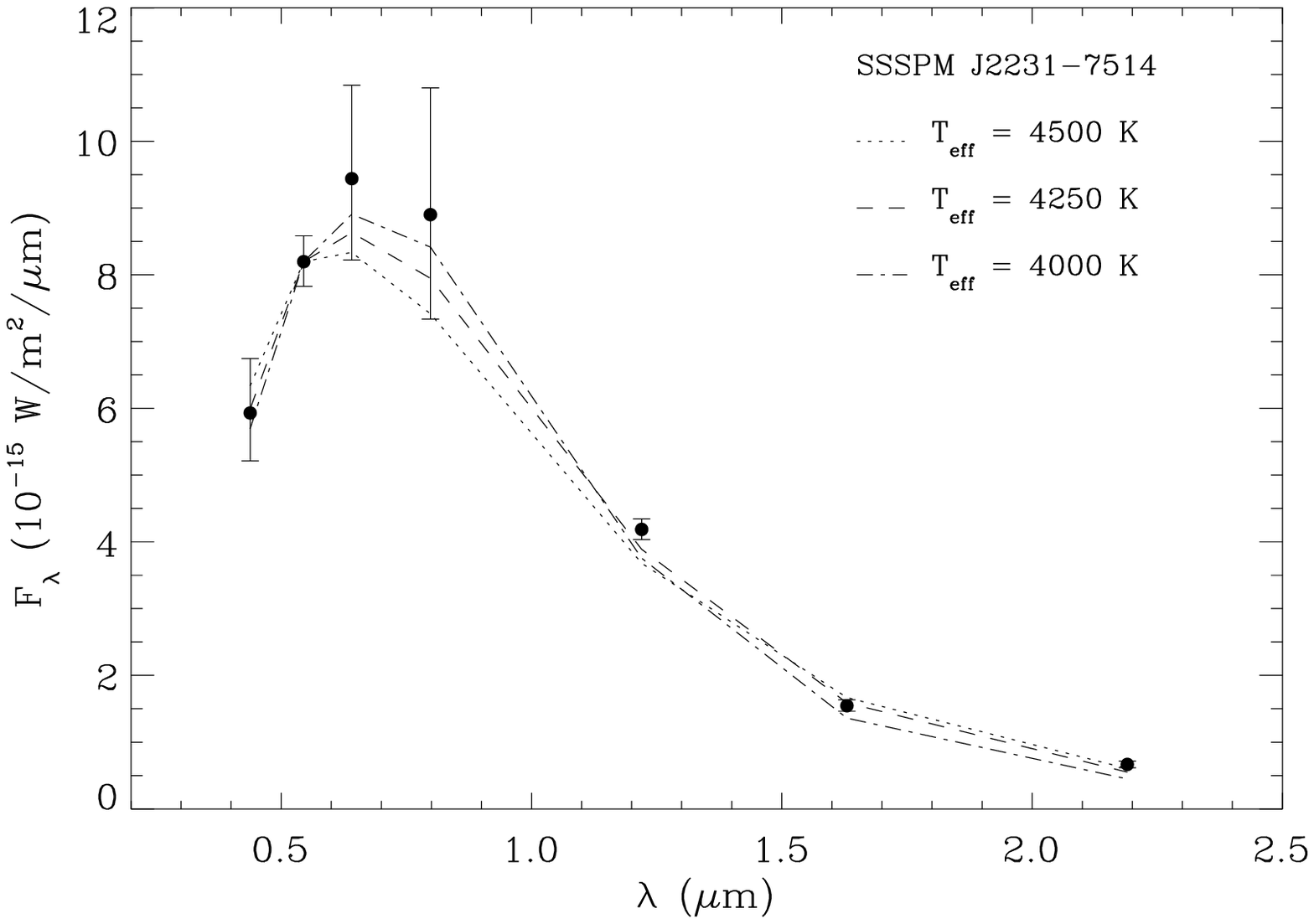}
\caption{Cool hydrogen atmosphere model fits to the
spectral energy distribution of SSSPM J2231$-$7514, assuming
log $g=8.0$ (\S2.3).
\label{fig1}}
\end{figure}

\begin{figure}

\plotone{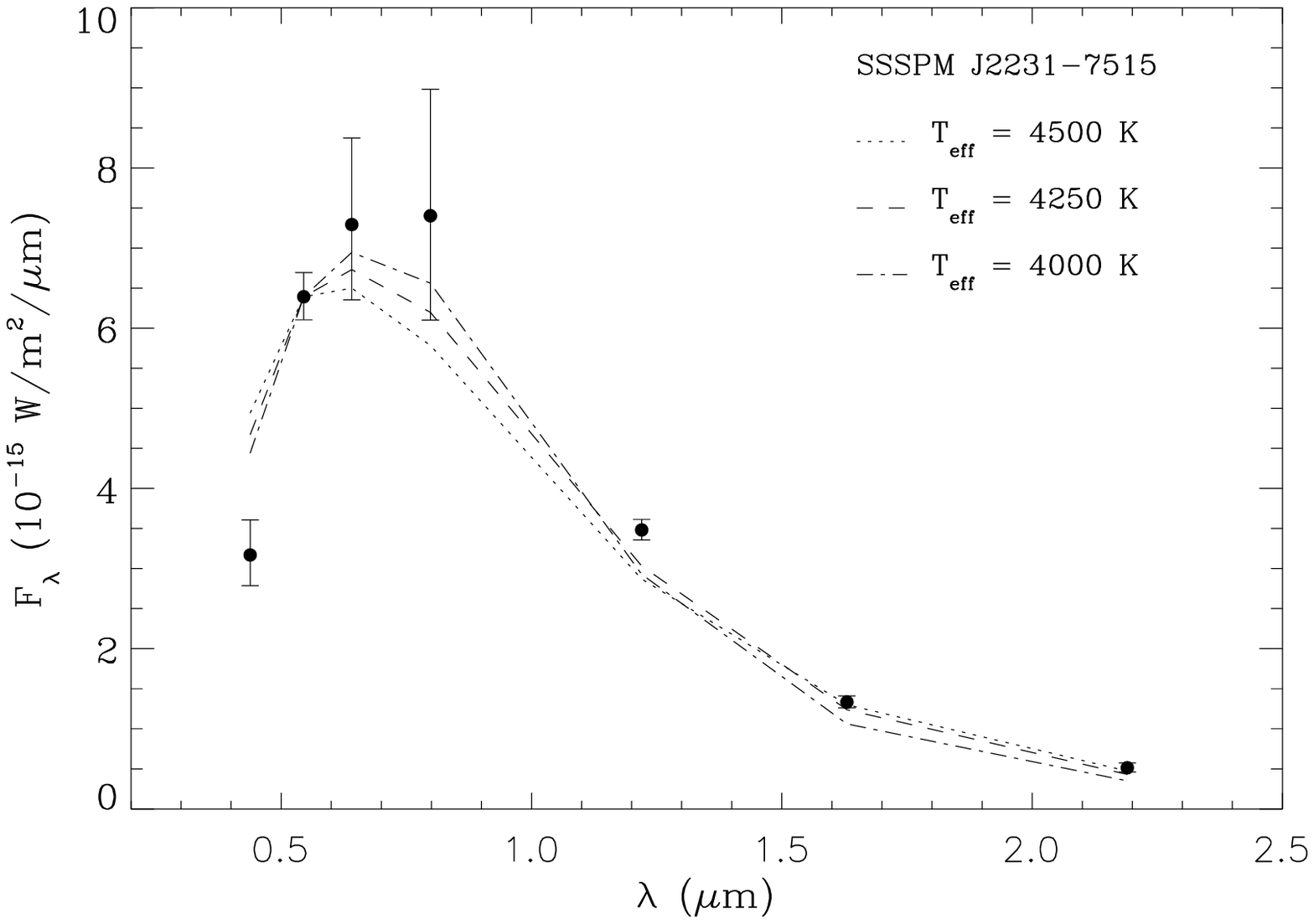}
\caption{Cool hydrogen atmosphere model fits to the
spectral energy distribution of SSSPM J2231$-$7515, assuming
log $g=8.0$ (\S2.3).
\label{fig2}}
\end{figure}

\begin{figure}

\plotone{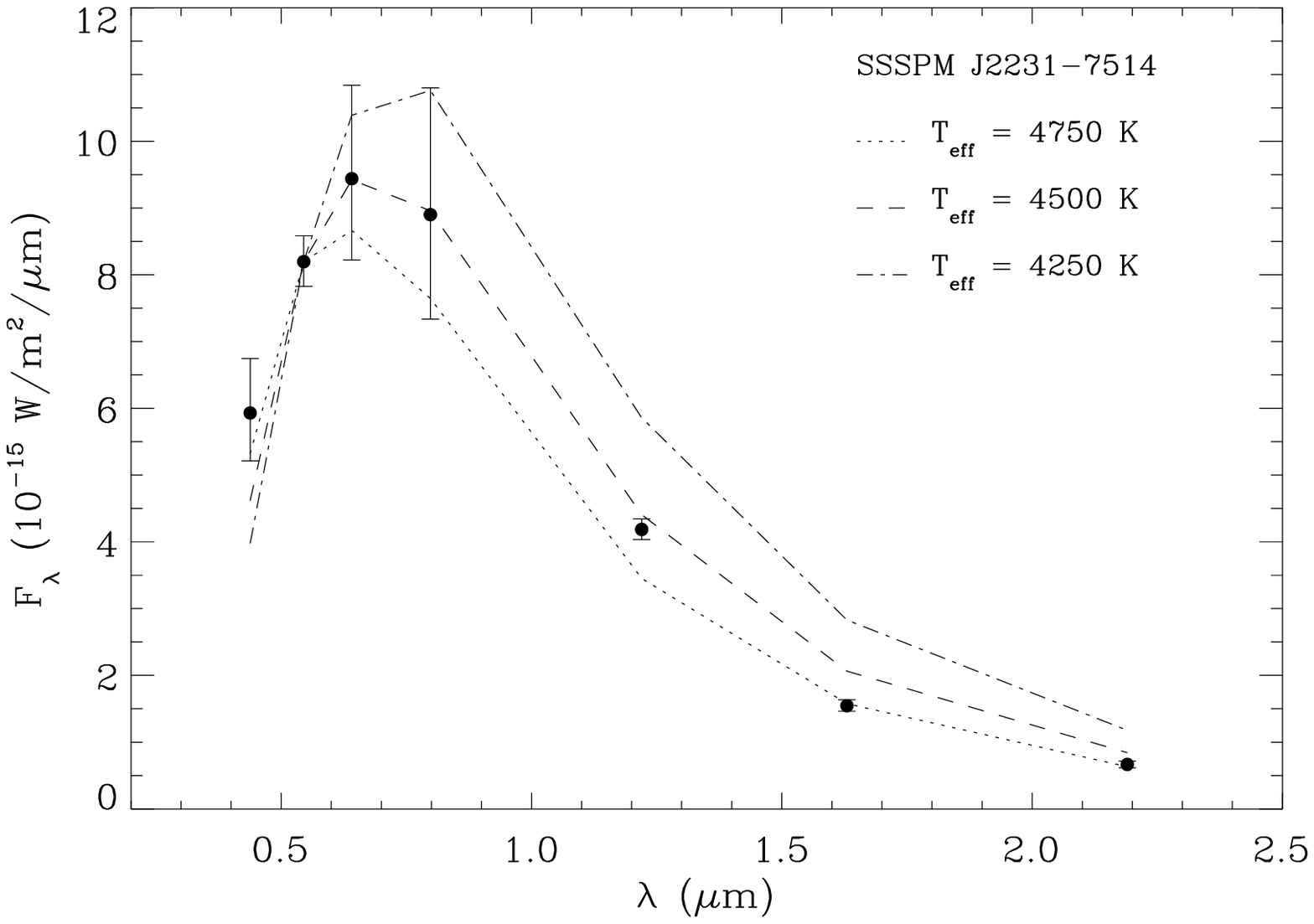}
\caption{Cool helium atmosphere model fits to the
spectral energy distribution of SSSPM J2231$-$7514, assuming
log $g=8.0$ (\S2.3).
\label{fig3}}
\end{figure}

\begin{figure}

\plotone{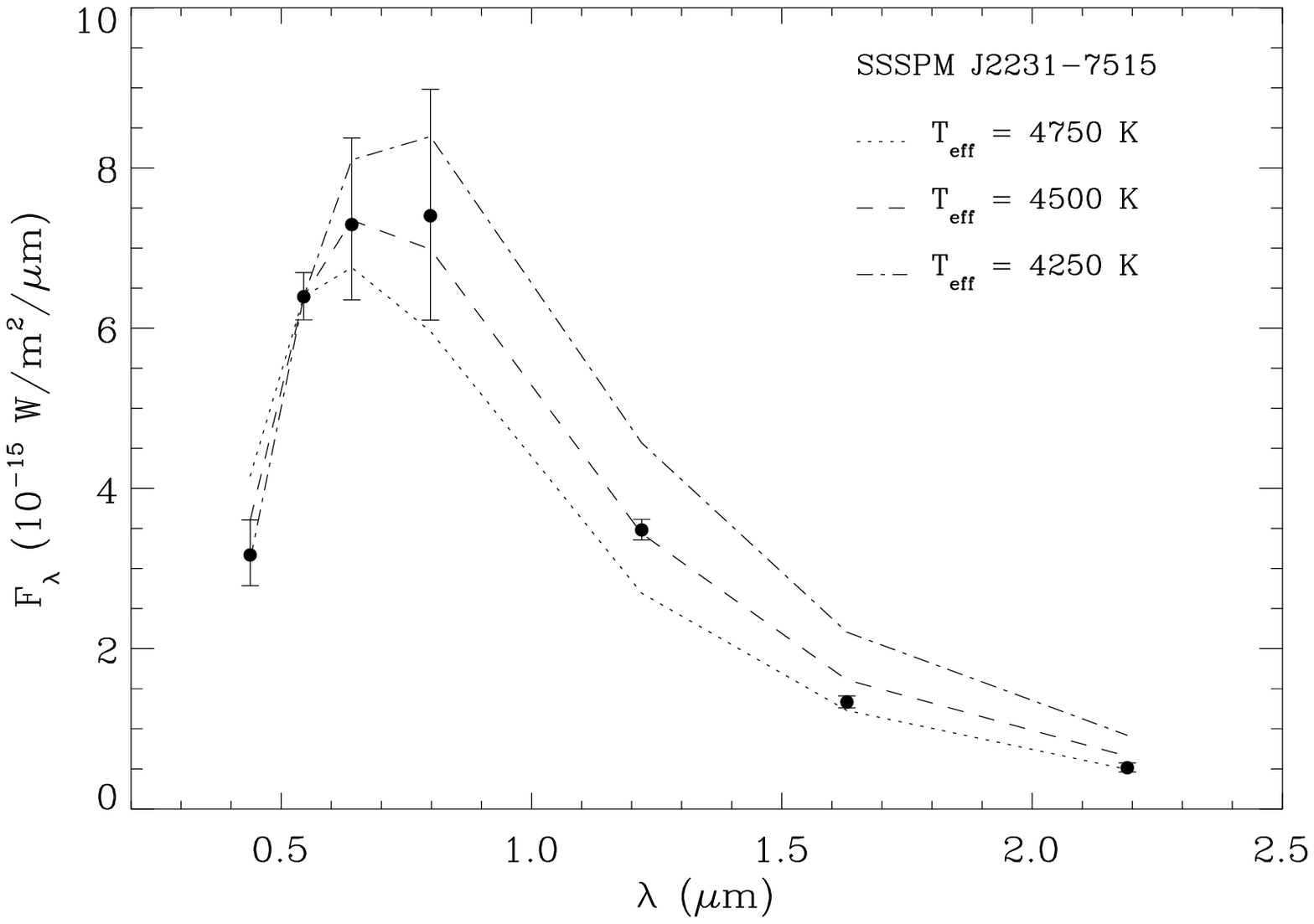}
\caption{Cool helium atmosphere model fits to the
spectral energy distribution of SSSPM J2231$-$7515, assuming
log $g=8.0$ (\S2.3).
\label{fig4}}
\end{figure}

$BVRIJHK$ magnitudes were converted to average fluxes following
the method of \citet{ber97} and fitted with the pure hydrogen and
helium model grids of P. Bergron (\citealt{ber95,ber95b}; 2002,
private communication).  A surface gravity of log $g=8.0$ was
assumed since the distance to the stars is not known.  The fits
are shown in Figures \ref{fig1}--\ref{fig4}.

The large error bars at $BRI$ are associated with the
external calibration of SSS photographic magnitudes.  These
errors might actually be underestimated here due to both error
propagation during the transformation to Johnson-Cousins $BRI$
and because the external errors reported in \citet{ham01b} were 
determined only for a small number of stars on plates in the
equitorial zone.  An illustration of the potential problem is
the fact that both SSSPM J2231$-$7514 and SSSPM J2231$-$7515 have
similar colors in all indices with the exception of $B-V$ where
they are different by 0.4 mag (a remnant from the original
photographic $B_J$).  This discrepancy is almost certainly due
to inaccuracies and a more conservative estimate of the errors
is 0.3 mag \citep{ham01b}.  This is an important consideration
when comparing the model predicted and measured fluxes at these
wavelengths.  One way to deal with these large uncertainties at
$BRI$ is to essentially ignore those data.  Another would be to
treat all data points equally, regardless of error.  A decent
compromise seems to be to give more weight to the $VJHK$ data, 
while still using all the available data in the fit.

The resulting preliminary spectral energry distributions
of both white dwarfs are matched quite well by $T_{\rm eff}=4250$
K pure hydrogen models.  Whereas the flux of the brighter component
in Figure \ref{fig1} is not inconsistent with the $T_{\rm eff}=4500$
K model, the flux of the the fainter component in Figure \ref{fig2} 
appears less agreeable with the higher temperature hydrogen model.
The flux estimates for both stars do not show good agreement with
$T_{\rm eff}<4250$ K hydrogen models, where significant CIA begins
to suppress near infrared flux and all infrared colors become
negative.  Mixed H/He atmosphere models predict even more CIA
for a given temperature and hence are also inappropriate
\citep{ber01a,opp01a}.  If all the data points are weighted
equally, then a pure helium model is applicable, yielding
$T_{\rm eff}\approx4500$ K for both stars (Figures \ref{fig3}
\& \ref{fig4}).

The fact that the data on both stars agree quite well with models
of the same $T_{\rm eff}$ does not contradict their measured magnitude
difference at $V$.  This difference could be due to their relative 
sizes (hence their mass ratio, which is assumed to be unity here).
A $0.1-0.2$ difference in log $g$ could explain their $\Delta V$
as could a $\sim200$ K difference in $T_{\rm eff}$ \citep{ber95b}.

\section{DISCUSSION}

\subsection{Ultracool White Dwarf Candidates}

There are only 4 white dwarfs with published optical and
near infrared data supporting their status as $T_{\rm eff}
<4000$ K degenerates.  These ultracool white dwarfs are, in
order of their discovery: LHS 3250, WD 0346+246, SDSS 1337+00,
GD392B \citep{har99,hod00,har01,far04}.  In addition, there are
several white dwarfs with published optical data that span the range
from candidate to all but certain ultracool white dwarfs.  These
are CE 51, F351-50, LHS 1402, WD 2356$-$209 \citep{iba00,opp01a,
opp01b,rui01,sal04}, and the five new Sloan stars recently reported
by \citet{gat04}.

Near infrared photometry indicates the proper motion selected
white dwarfs SSSPM J2231$-$7514 and SSSPM J2231$-$7515 both have
$T_{\rm eff}\ga4250$ K.  For log $g\sim8.0$, this would put the
wide binary at a distance of around 15 pc, assuming 4500 K for
the brighter and 4250 K for the fainter component.  They may be
the coolest white dwarfs known within 20 pc.  There are only two
white dwarfs with measured $\pi>50$ mas and $T_{\rm eff}<4500$ K
as determined by full spectroscopic and photometric analyses
including near infrared data; LHS 239 \& ER 8 \citep{ber97,
ber01b,hol02}.

As \citet{ber03} points out, the spectral energy distributions
of cool white dwarfs are not well constrained by optical data alone.
Colors such as $V-I$ reach a maximum redness and then become bluer
again due to CIA -- yielding two possible temperatures for a given
value of $V-I$ \citep{ber03}.  Hence any white dwarf study claiming
sub 4000 K temperatures should present the requisite near infrared data.

Optical spectroscopy also has pitfalls.  Blackbody fits to the
$4300-6800$ \AA{} flux calibrated spectra of SSSPM J2231$-$7514
and SSSPM J2231$-$7515 yielded temperatures of 3810 K and 3600 K, 
respectively \citep{sch02}.  The analysis here shows these
temperatures are likely to be underestimated by at least 650 K.
In contrast, the blackbody fits to the $4000-8500$ \AA{} flux calibrated
spectrum of WD 0346+246 yielded temperatures $100-150$ K {\it higher}
than $T_{\rm eff}=3750$ K as determined by parallax and total 
integrated flux \citep{ham97,hod00,opp01a}.  This could be because
white dwarfs with significant CIA in the near infrared will have
some of their flux redistributed toward higher energies.  Assuming
the flux calibration of \citet{sch02} is correct, blackbodies simply
do not provide a good estimate of $T_{\rm eff}$ for cool white dwarfs.

The flux calibrated optical spectra of F351-50 and
F821-07 (LHS 542) were fitted with 3500 K and 4100 K
blackbodies respectively \citep{iba00}.  F351-50 was noted to
have ``a substantial depression of the flux redward of 6500
\AA{}... precisely as was originally seen in WD 0346+246'',
while LHS 542 is noted as having ``a similar spectral shape
to  WD 0346+246'' \citep{iba00}.  First, WD 0346+246 does not
show flux suppression in the optical but approximates a
$T\approx3900$ K blackbody fairly well out to $\sim9000$ \AA{}
\citep{ham97,hod00,opp01a}.  Second, the most reliable effective
temperature determination of LHS 542 is 4720 K, based on its
trigonometric parallax plus optical and near infrared photometry
\citep{leg98,ber03}.  There is certainly no flux deficit out to
$2.2\mu$m as seen in the measured data and model fit shown in
Figure 2 of \citet{ber03} for LHS 542.  Third, there is no
corroborating evidence of a flux deficit in F351-50.  Its optical
spectrum as shown in Figure 3 of \citet{opp01a} appears to have a
flatter slope than WD 0346+246 out to 10,000 \AA{} and looks fairly
consistent with the 4000 K blackbody plotted in the same Figure.
Hence there appears to be a problem in either the flux calibration
or the blackbody in Figure 1 of \citet{iba00} that causes both white
dwarfs to appear cooler.  The 620 K difference in the effective
temperatures reported for LHS 542 by \citet{iba00} and \citet{ber03},
if added to the 3500 K temperature estimate for F351-50, yields 4120
K -- exactly the value obtained by \citet{ber03} as one of two
possibilities for F351-50 based on optical data alone.  Additional
data has confirmed this higher temperature as likely (P. Bergeron
2004, private communication).

\subsection{The Origin of Ultracool Degenerates}

An important goal is to understand the origin of ultracool white
dwarfs, both in the disk and the halo.  Halo white dwarfs can be older
than 10 Gyr and have therefore, according to models, had enough time to
cool to sub 4000 K temperatures, regardless of atmospheric composition
and mass \citep{ber95,han99}.  Normal mass ($M\approx0.6$ $M_{\odot}$)
disk white dwarfs on the other hand, generally have not had enough time
to attain ultracool temperatures with the exception of very low mass
($M\leq0.4$ $M_{\odot}$) or very high mass ($M\geq1.0$ $M_{\odot}$)
cases \citep{ber95,han99,ser01}.

So far, there is both solid and tentative evidence for ultracool disk
white dwarfs of low mass \citep{har01,far04}.  These remnants are likely
to be the products of close binary evolution rather than single stars
evolved from the main sequence \citep{mar95}.  Possibly awaiting detection
are the much fainter high mass ultracool white dwarf counterparts
\citep{rui95,far04}.  Trigonometric parallax measurements will tell
if any of the new Sloan ultracool white dwarfs are massive \citep{gat04}.
The differential cooling between low, normal and high mass degenerates may
be the most important reason to distinguish between white dwarfs warmer or
cooler than $\sim4000$ K.  Specifically, cool and ultracool disk white
dwarfs may have separate formation channels.  

Given the fact that the peak flux for ultracool white dwarfs is
in the optical region of the spectrum, the dearth of detections may
be telling.  However, the available data on the coolest degenerates is
a product of the finite age of the local disk convolved with its star
formation history plus the ability of various searches to identify
them.  Astronomers must first be confident of their ability to detect
them before understanding their relative numbers, origins, and overall
astrophysical implications.

\subsection{Classification of CIA White Dwarfs}

Spectrally distinct stars should be classified distinctly.  However,
spectral assignment must depend on observed features only and be model
independent.  In the accepted scheme of \citet{mcc99} for white dwarfs, 
the effective temperature index is completely independent of spectral
type.  Therefore, any designation for white dwarfs displaying CIA would
be independent from effective temperature.

Technically speaking, are white dwarfs with CIA featureless?
Although potentially an extremely broad feature in pure hydrogen
atmospheres, CIA is essentially a continuum opacity in all white dwarfs
for which it has been observed.  This opacity is virtually undetectable
until very strong, where it is evident in flux calibrated optical or
near infrared spectra \citep{har99,hod00,har01,gat04}.  Therefore ``DC''
alone may not be the most appropriate designation for these degenerates
(this is especially true in light of the possibility that pure helium
atmosphere stars cooler than 4000 K may exist and await discovery).

Interestingly, with the exception of the DQ9.5 star LHS 1126
($T_{\rm eff}=5400$ K, \citealt{ber94}), there are currently no
other cool white dwarfs at temperatures significantly above 4000
K with significant CIA as evidenced by blue near infrared colors.
All other white dwarfs with CIA are currently suspected to be
DC13+ stars.

\section{CONCLUSION}

An analysis of existing data on SSSPM J2231$-$7514 and SSSPM
J2231$-$7515 indicates $T_{\rm eff}\approx4250$ K for both white
dwarfs.  This value should be considered preliminary as higher
signal to noise optical and near infrared photometry is needed.
If the 2MASS data are accurate, the near infrared colors of these
white dwarfs are red and not consistent with significant flux
suprression due to CIA.  These two stars, among others, may 
represent the coolest effective temperatures attainable by normal
mass, single white dwarf evolution in the disk of the Galaxy.
Degenerates with temperatures below 4000 K, the ultracool
degenerates, may be the unique signature of halo white dwarfs
and disk white dwarfs of atypical mass.

\acknowledgments

Some data used in this paper are part of the Two Micron All
Sky Survey, a joint project of the University of Massachusetts
and the Infrared Processing and Analysis Center (IPAC)/CIT,
funded by NASA and the National Science Foundation (NSF).
2MASS data were retrieved from the NASA/IPAC Infrared Science
Archive, which is operated by the Jet Propulsion Laboratory,
CIT, under contract with NASA.  The author acknowledges the
Space Telescope Science Insititute for use of the digitized
version of the POSS I \& II plates.  This research has been
supported in part by grants from NASA to UCLA.

\clearpage

\clearpage

\clearpage

\clearpage

\clearpage

\begin{deluxetable}{lccc}
\tablecaption{Optical \& Near Infrared Photometric Data\label{tbl-1}}
\tablewidth{0pt}
\tablehead{
\colhead{Band} 				&
\colhead{$\lambda_{0}(\mu$m)} 		&
\colhead{SSSPM J2231$-$7514}		&
\colhead{SSSPM J2231$-$7515}}

\startdata

$B$		&0.44	&$17.56\pm0.14$	&$18.24\pm0.14$	\\
$V$		&0.55	&$16.60\pm0.05$	&$16.87\pm0.05$ \\
$R$ 		&0.64	&$15.89\pm0.15$	&$16.18\pm0.15$	\\
$I$ 		&0.80	&$15.25\pm0.21$	&$15.45\pm0.21$	\\
$J$		&1.25	&$14.66\pm0.04$	&$14.86\pm0.04$	\\
$H$ 		&1.63	&$14.66\pm0.06$	&$14.82\pm0.06$	\\
$K_s$		&2.16	&$14.44\pm0.08$	&$14.72\pm0.12$ \\

\enddata

\tablecomments{Near infrared data all have ${\rm SNR}>10$ and
are taken from 2MASS \citep{cut03}.  Optical data are taken from
\citet{sch02} with $BRI$ converted from photographic magnitudes
to the Johnson-Cousins system.  The errors in $BRI$ are from
\citet{ham01b} and do not include any conversion errors.}

\end{deluxetable}

\end{document}